\documentclass[12pt]{iopart}
\usepackage{iopams}
\usepackage[dvips]{epsfig}
\usepackage{color}
\newcommand{\be}{\begin{equation}}
\newcommand{\ee}{\end{equation}}
\newcommand{\bea}{\begin{eqnarray}}
\newcommand{\eea}{\end{eqnarray}}

\newcommand{\smallfig}{0.45\textwidth}

\begin {document}

\title{Spin Ladder with Anisotropic Ferromagnetic Legs in a Transverse Magnetic Field }
\author{G. I. Japaridze$^1$, A. Langari$^{2,3}$, and S. Mahdavifar$^4$  }

\address{$^1$ International Center for Condensed matter Physics,
Universidade de Brasilia, 70904-910 Brasilia-DF, Brazil \, and  \,
Andronikashvili Institute of Physics, Tamarashvili str. 6, 0177,
Tbilisi, Georgia}
\address{$^2$ Physics Department, Sharif University of Technology, Tehran 11365-9161, Iran}
\address{$^3$ Institute for Studies in Theoretical Physics and Mathematics (IPM), Tehran 19395-5531, Iran}
\address{$^4$ Institute for Advanced Studies in Basic Sciences, Zanjan 45195-1159, Iran}

\begin{abstract}

We study the ground state phase diagram of a two-leg spin
ladder with anisotropic ferromagnetic leg couplings under the influence of a
symmetry breaking transverse magnetic field by the exact
diagonalization technique. In the case of antiferromagnetic coupling
between legs we identified two phase transitions in the plane of
magnetic field vs interchain coupling strength. The first one
corresponds to the transition from the gapped rung-singlet phase to
the gapped stripe-ferromagnetic phase. The second one represents the
transition from the gapped stripe-ferromagnetic phase into the fully
polarized ferromagnetic phase.

\end{abstract}
\maketitle

\section{Introduction}\label{intro}

There has been recently considerable interest in study of the
magnetic field-induced effects in low-dimensional quantum spin
systems. In particular, critical properties of the spin $S=1/2$
{\em isotropic antiferromagnetic} two-leg ladders in a magnetic
field have been a field of intense studies. This seems pertinent
in the face of great progress made within the last  years in fabrication of
such ladder compounds \cite{RiceDagotto}. Moreover, since the
antiferromagnetic two-legs ladder systems have a gap in the spin
excitation spectrum, they reveal extremely rich quantum behavior
in the presence of magnetic field (See for recent review [2]). Such
quantum phase transitions in spin systems with gapped excitation spectrum
were indeed studied experimentally [3-8]. On the theoretical side
these transitions were intensively investigated using different
analytical and numerical techniques [9-23].

\begin{figure}[tbp]
\centerline{\includegraphics[width=0.45\columnwidth]{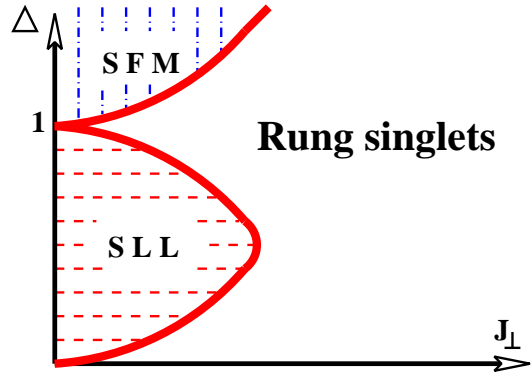}}
\caption{Schematic picture of the ground state phase diagram of
the two-leg ladder as a function of the intraleg exchange
anisotropy ($\Delta$) and isotropic interleg coupling
($J_{\perp}$) in the considered in this paper sector of the phase
diagram corresponding to the competing ferromagnetic intraleg
($\Delta>0$) and interleg antiferromagnetic ($J_{\perp} > 0$)
exchange. The Spin-Luttinger-Liquid phase is denoted by SLL, while
the Stripe-Ferromagnetic phase - by SFM. (From Ref \cite{VJM_1}.)}
\label{Fig:FL_3_Fig_1}
\end{figure}

Ladder systems with {\em ferromagnetic legs} are much less studied.
Partly, the reason is that spin-ladders
with ferromagnetic legs are not still fabricated. However, from
the theoretical point of view these systems are extremely
interesting, since they open a new wide polygon for the study of
complicated quantum behavior, unsuspected in more conventional
spin systems [9,24-30]. The variety of open possibilities is
clearly seen from the weak-coupling phase diagram of a two-leg
ferromagnetic ladder which is a function of the intraleg exchange
anisotropy ($\Delta$) and the isotropic interleg coupling
($J_{\perp}$) \cite{VJM_1}, it is presented in
Fig.\ref{Fig:FL_3_Fig_1}.  In addition to the fully gapped rung-singlet phase
(RS) (common in the case of antiferromagnetic ladder
\cite{Schulz1,SNT_96}), the ground state
phase diagram contains the gapless Spin-Luttinger-Liquid phase with
the easy-plane anisotropy (SLL) and the stripe-ferromagnetic (SFM)
phase which are realized only in the case of ferromagnetic legs
($\Delta >0$)\cite{Schulz1,VJM_1}.

The very rich and interesting new phenomena arise in the case of
competing ferromagnetic intraleg and antiferromagnetic interleg
couplings, in particular in the presence of symmetry breaking
magnetic field \cite{VJM_1,VJM_2}. The effect of the uniform
magnetic field, applied parallel to $Z$ (quantization) axes, on
the properties of the two-leg ladder systems was first studied (in
the case of equivalent spin $S=1$ Heisenberg chain model) by H.
Schulz \cite{Schulz1}. In the case of gapped rung-singlet phase
the magnetization appears only at finite critical value of the
magnetic field equal to the spin gap
\cite{Schulz1,ChitraGiamarchi}. This behavior is generic for the
spin gapped $U(1)$ symmetric systems in a magnetic field which
leaves the in-plane rotational invariance unchanged \cite{JNW} and
belongs to the universality class of the
commensurate-incommensurate (C-IC) transitions \cite{JN_PTr}.

The effect of an {\em uniform transverse
field} in the case of  $U(1)$ symmetric phase is highly nontrivial. In the
case of classical spin chains this effect has been studied already
a decade ago \cite{Mikeska}. However, in the case of quantum
antiferromagnetic $XXZ$ chain this problem is still the subject of
intense recent studies [32-41].

In this paper, we address the similar problem and study the effect
of {\em uniform transverse} magnetic field on the ground state
phase diagram of a two-leg ladder with {\em anisotropic,
ferromagnetically interacting} legs coupled by antiferromagnetic
interleg exchange. The Hamiltonian of the model under
consideration is given by
\begin{eqnarray}
H=H_{leg}^{(1)} + H_{leg}^{(2)} + H_{\bot}
\label{Hamiltonian}
\end{eqnarray}
where the Hamiltonian for leg $"j"$ is
\begin{eqnarray}
H_{leg}^{(j)} = - J
\sum_{n=1}^{N}\big[\left(s_{j,n}^{x}s_{j,n+1}^{x}+s_{j,n}^{y}s_{j,n+1}^{y}+\Delta
s_{j,n}^{z}s_{j,n+1}^{z}\right) + \mathrm{h^{ext}}
s_{j,n}^{x}\big]
\end{eqnarray}
and the interleg coupling is
\begin{eqnarray}
H_{\bot} = J_{\bot}\sum_{j=1}^{N} \left( s_{1,j}^{x}s_{2,j}^{x}+s_{1,j}^{y}s_{2,j}^{y}+
s_{1,j}^{z}s_{2,j}^{z} \right) \, .
\end{eqnarray}
Here $s_{j,n}^{x, y, z}$ are spin $s=1/2$ operators on the $n$-th
rung, and the index $j=1,2$ denotes the ladder legs. The intraleg
coupling constant is ferromagnetic, $J>0$, and therefore the
limiting case of isotropic ferromagnetic legs corresponds to
$\Delta=1$ and the external magnetic field is proportional to $h= J \mathrm{h}^{ext}$.

The model (\ref{Hamiltonian}) has been studied recently using the
continuum-limit bosonization approach \cite{VJM_1,VJM_2}. The main
attention in this studies was focused on the investigation of new
field induced effects in the sector of phase diagram corresponding
to the gapless easy-plane phase, present in the case of weak
antiferromagnetic intraleg exchange $0<J_{\bot}<J$
\cite{Schulz1,VJM_1}. It has been shown, that in the presence of
an infinitesimally small transverse magnetic field which breaks
the in-plane rotational symmetry the gapless phase is unstable
towards the gapped stripe-ferromagnetic (SFM)
phase \cite{VJM_2}. The SFM phase is characterized by the uniform
magnetization in the direction of applied field and opposite
magnetization of legs in the in-plane direction perpendicular to
the field. It has been also shown, that when the magnetic field
exceeds some critical value, the interleg antiferromagnetic order
disappear and the system passes into the fully polarized
ferromagnetic phase \cite{VJM_2}.

\begin{figure}[tbp]
\centerline{\includegraphics[width=0.85\columnwidth]{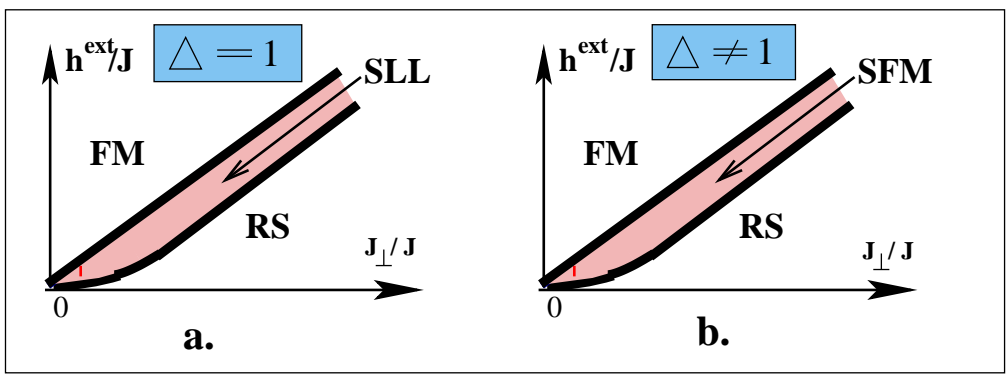}}
\caption{Qualitative sketch of the weak-coupling phase diagram of
a two-leg ferromagnetic ladder as a function of the applied
magnetic field ($\mathrm{h}^{ext}$) and the isotropic interleg
coupling ($J_{\perp}$) \cite{VJM_2}. In the case of isotropic
spin-exchange interaction in legs ($\Delta=1$) the spin gapped
Rung-Singlet (RS) phase is separated from the fully polarized
ferromagnetic (FM) phase via the gapless Spin-Luttinger-Liquid
(SLL) phase. In the case of finite intraleg exchange anisotropy
($\Delta \neq 1$) the RS phase is splitted from the FM phase by
the gapped Stripe-Ferromagnetic (SFM) phase. }
\label{Fig:FL_3_Fig_2}
\end{figure}

In the opposite case of strong rung exchange $J_{\bot}\gg J$ an
analytical description of the magnetic properties of the system in
the presence of transverse magnetic field is available only in the
$SU(2)$ symmetric case at $\Delta=1$. In this limiting case it has
been shown that the isotropic ladder in a magnetic field shows two
second order phase transitions: at
$\mathrm{h}^{ext}=\mathrm{h}^{ext}_{c1}$ from a spin gapped
rung-singlet (RS) phase to a gapless Spin-Luttinger-Liquid (SLL)
phase and at $\mathrm{h}^{ext}=\mathrm{h}^{ext}_{c2}$ a transition
from a SLL phase into the fully polarized ferromagnetic (FM) phase
(see Fig.\ref{Fig:FL_3_Fig_2} a) \cite{VJM_2}.

In the case of finite intraleg exchange anisotropy ($\Delta \neq
1$), when the effect of symmetry breaking  transverse field is
important, an analytical solution is not available. However,
based on the qualitative estimations and symmetry analysis it has
been shown that the two-step transition from the RS into the FM
phase is present also in the case of ladder with anisotropic legs
\cite{VJM_2}. In the same work it has been also predicted that for
$\Delta \neq 1$ the RS phase is separated from the FM phase by the
gapped SFM phase (see Fig.\ref{Fig:FL_3_Fig_2} b).

In this paper we continue our studies of the anisotropic spin
ladder in the transverse magnetic field. In particular we apply
the modified Lanczos method to diagonalize numerically finite $(N
= 12, 16, 20, 24)$ ladder systems. Using the exact
diagonalization results, we calculate the spin gap, magnetization,
various spin-structure factors and the rung dimerization order
parameter as a function of applied transverse field for various
values of the anisotropy parameter $\Delta \neq 1$. Based on the
exact diagonalization results we obtain the magnetic phase diagram
of the ferromagnetic ladders with anisotropic legs showing two
phase transitions in the plane of magnetic field vs interchain
coupling in agreement with the predictions made in Ref. [27].

The outline of the paper is as follows: In section II we briefly
discuss the model in the strong rung coupling limit and derive the
effective spin chain Hamiltonian to outline the symmetry aspects
of the problem. In section III we present the results of  exact
diagonalization calculations using the modified Lanczos method.
Finally we conclude and summarize our results in section IV.


\section{Large rung coupling results}

In this section we briefly discuss the model (\ref{Hamiltonian})
in the limiting case of strong rung coupling $J_{\bot}\gg J$. In
this limit the model (1) can be mapped onto the effective
spin-chain Hamiltonian \cite{Chaboussant2,Mila}. This allows to outline
the symmetry aspects of the problem under consideration.

At $J_{\bot}\gg J$ it is convenient to discuss the model by
representing the {\em site}-spin algebra in terms of {\em
on-bond}-spin operators \cite{SachdevBhatt}. Indeed an individual
rung may be in a singlet or a triplet state with corresponding
spectrum given by
$$
E_{\pm}=\left({J_{\bot}\over 4}\ \pm h^{ext}\right),\,\,
E_0={J_{\bot}\over 4},\,\, E_s=-{3J_{\bot}\over 4}.
$$
At $h^{ext} \leq J_{\bot}$, one component of the triplet becomes
closer to the singlet ground state such that for a strong enough
magnetic field we have a situation that the singlet and the $S_{z} = -1$
component of triplet create a new effective spin $\tau=1/2$
system. One can easily project the original ladder Hamiltonian
(\ref{Hamiltonian}) on the new singlet-triplet subspace
$$
|\Uparrow\rangle  \equiv |s\rangle  =
\frac1{\sqrt2}[|\uparrow\downarrow\rangle - |\downarrow\uparrow\rangle]\, , \qquad
|\Downarrow\rangle  \equiv  |t^{-}\rangle  =
|\downarrow\downarrow\rangle \, .
\nonumber
$$
This leads to the definition of the effective spin 1/2 operators
\begin{equation}
S^+_{n,\alpha=1,2} =  (-1)^{n+\alpha} \frac1{\sqrt2}\tau^{+}_{n}\, , \qquad
S^z_{n,\alpha=1,2} = \frac{1}{4} [I + 2\tau^{z}_{n}]
\label{tau}
\end{equation}
The effective Hamiltonian  in terms of the effective spin operators
(\ref{tau}) up to the accuracy of an irrelevant constant  becomes the Hamiltonian of the
spin 1/2 {\em fully anisotropic} $XYZ$ chain in an effective
magnetic field
\be H_{\emph{eff}}  = -J \sum_i
[\frac{1}{2}\tau^{z}_{n}\tau^{z}_{n+1} + \tau^{y}_{n}
\tau^{y}_{n+1} +\Delta \tau^{x}_{n} \tau^{x}_{n+1}] +
\mathrm{h}^{eff} \sum_n \tau^{z}_{n}\, ,
\label{EffectiveHamiltonian2} \ee
where $\mathrm{h}^{eff} = \mathrm{h}^{ext} - J_\perp + J/2$. Note that in deriving (\ref{EffectiveHamiltonian2}), we have used the rotation in the effective spin space
which interchanges the $x$ and $z$ axes.

At $\Delta = 1$, the effective problem reduces to the theory of the
$XXZ$ chain with a \emph{fixed ferromagnetic} $XY$ anisotropy of
$1/2$ in a magnetic field, which allows for rigorous analysis
\cite{VJM_2}. The gapped phase at $\mathrm{h}^{eff}  <
\mathrm{h}^{eff} _{c1}$ for the ladder corresponds to the negatively
saturated magnetization phase for the effective spin chain, whereas
the massless phase for the ladder corresponds to the finite
magnetization phase of the effective spin-1/2 chain. The critical
field $\mathrm{h}^{eff}_{c2}$
corresponds to the fully magnetized phase of the effective spin
chain where the ladder is totally magnetized. From the exact ground state phase diagram of the anisotropic
$XXZ$ chain in a magnetic field \cite{Takahashi}, it is easy to
check that $\mathrm{h}^{eff}_{c1,c2}= \mp J/2$.
For the
{\em isotropic ferromagnetic ladder} in a magnetic field
this corresponds to
a transitions from rung-dimers to SLL phase at $\mathrm{h}^{ext}_{c1}=
J_{\bot}-J$ and a transition from  SLL phase into the fully
polarized phase at $\mathrm{h}^{ext}_{c2}=J_{\bot}$ (see
Fig.\ref{Fig:FL_3_Fig_2} a). In this paper we compare this values
of critical fields
with the numerical results of exact diagonalization method.

Away from the isotropic point $\Delta =1$ the effective
Hamiltonian (\ref{EffectiveHamiltonian2}) describes the fully
anisotropic ferromagnetic $XYZ$, chain in a magnetic field that is
directed perpendicular to the easy axis.  For the particular value
of magnetic field $\mathrm{h}_{eff}=0$, the effective $XYZ$ chain is long
range ordered in $Y-$direction \cite{McCoy}, accordingly the original
ladder system being ordered in the direction perpendicular to the
applied magnetic field with opposite magnetization on  legs
(stripe-ferromagnetic phase). For larger values of the effective
field it is clear that this SFM order will be replaced either by
the rung singlet phase or the phase with only one order parameter
- magnetization along the applied field.

The exact diagonalization study
of finite ladder systems (up to length $L=12 (N=24)$) show that the
magnetic phase diagram of the ferromagnetic ladder with
anisotropic legs in a transverse magnetic field really presents two
quantum phase transitions with increasing magnetic field. The
first transition is from a gapped RS phase to the gapped SFM phase
and the second one from a SFM phase into the fully
polarized FM phase (see Fig.\ref{Fig:FL_3_Fig_2}b).

\section{Exact diagonalization results}

In order to explore the nature of the spectrum and the phase
transition,  we use the modified Lanczos method \cite{Langari_05,ggrosso}
to diagonalize numerically finite ($N=12,16,20,24$ sites) ladder systems. The
energies of the few lowest eigenstates were obtained for chains with periodic boundary
conditions.


\begin{figure}[tbh]
\centerline{\includegraphics[width=\smallfig]{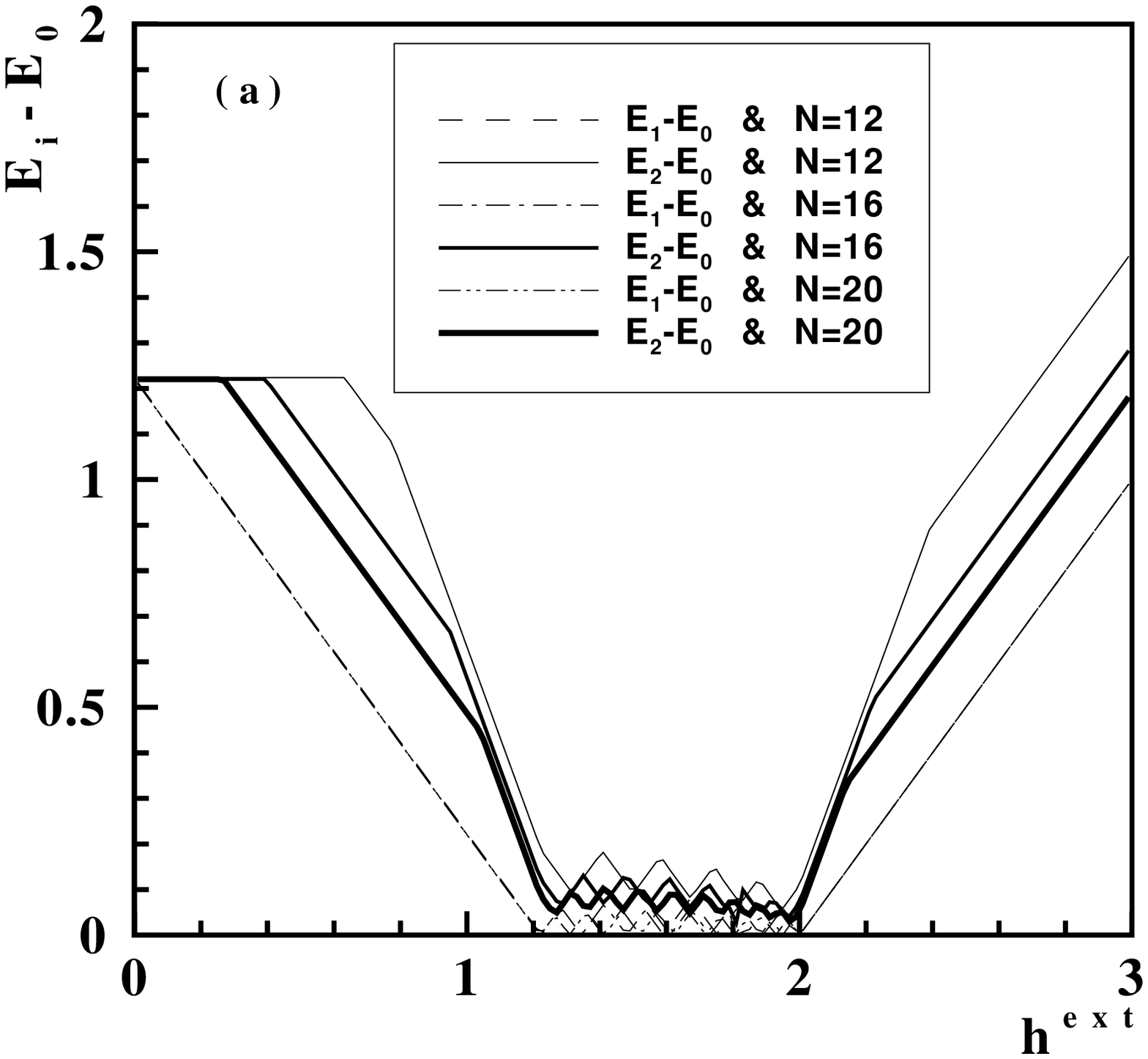}
\includegraphics[width=\smallfig]{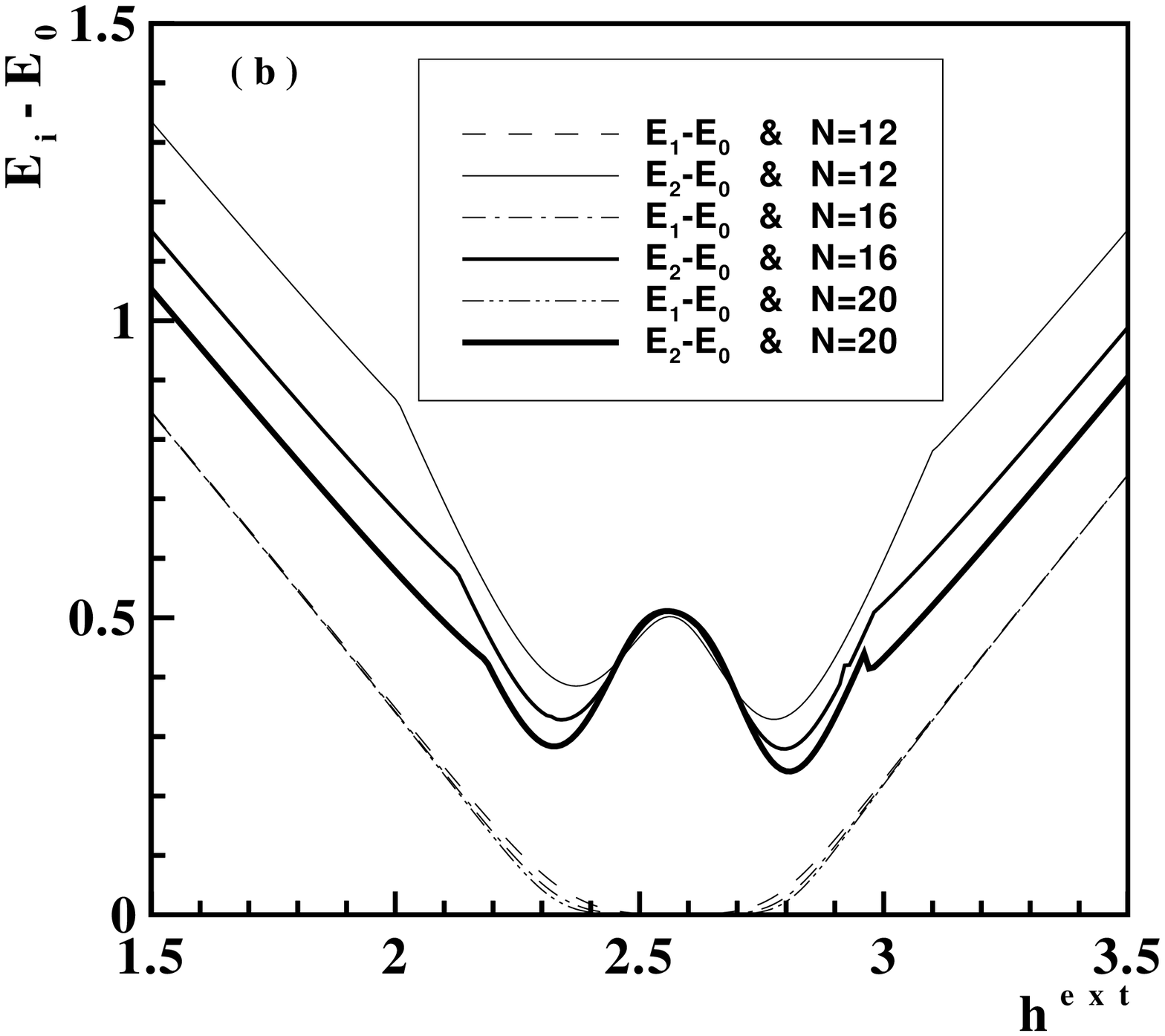}}
\caption[DIAGONAL]{Difference between the energy of the two lowest levels and
the ground state energy as a function of the magnetic field,
(a) $\mathrm{h}^{ext}$ for $J_{\bot}=2J$ and $\Delta=1.0$ and (b) for $J_{\bot}=3J$ and $\Delta=0.5$ including different ladder lengths $N=12,16,20$.}
\label{Fig:ChargeGaps}
\end{figure}


To show, that the transition lines can be easily observed from the
numerical calculations of small systems we start our consideration
by the case of {\em isotropic chains}.
In this case, the magnetic field term commutes with the spin-spin exchange terms which
makes the total x-component ($S^x_{tot}$) of spin a conserved quantity. Thus, the energy levels of ladder
are labeled by $S^x_{tot}$. We have computed the vlaue of energy levels in the absence of magnetic field and added
the effect of magnetic field by  $-J  \mathrm{h}^{ext} S^x_{tot}$. This makes very high accuracy for the value of energy levels.
In
Fig.\ref{Fig:ChargeGaps}a we have plotted the three lowest levels
of  $N=12,16,20$ ladder with $J_{\bot}=2J$ as a function
of the external field $\mathrm{h}^{ext}$. We determine the
excitation gap in the system as the difference between the first
excited state and the ground state. As it is clearly seen from
this figure in the case of zero magnetic field the spectrum of the
model is gapped. For $\mathrm{h}^{ext} \neq 0$ the gap decreases
linearly with $\mathrm{h}^{ext}$ and vanishes at the critical field, $\mathrm{h}^{ext}_{c1}$.
This is the first level crossing between the ground state energy and the first excited state one.
To get an accurate estimate of $\mathrm{h}^{ext}_{c1}$ we have obtained the first level crossing for
system sizes of $N=12,16, 20, 24$. The finite size behaviour of this values leads us to
$\mathrm{h}^{ext}_{c1}=1.2\pm0.01$ for $N\rightarrow \infty$. The spectrum remains gapless
for $\mathrm{h}^{ext}_{c1}< \mathrm{h}^{ext}
<\mathrm{h}^{ext}_{c2}$ and becomes once again gapped for $\mathrm{h}^{ext} > \mathrm{h}^{ext}_{c2} = 2.0$.
We got $\mathrm{h}^{ext}_{c2} = 2.0$ as an exact value since there was no finite size
correction at this value. We should mention that the critical field values obtained in the previous section
come from the first order (perturbation) effective Hamiltonian approach and are not exact values.
However, the values of the critical fields
$\mathrm{h}^{ext}_{c1}$ and $\mathrm{h}^{ext}_{c2}$ obtained from
studies of the finite chains are very close to the values get in
the previous section.
With increasing
field, for $\mathrm{h}^{ext} \gg J_{\bot}, J$ the gap increases
linearly with $\mathrm{h}^{ext}$.

It  can be seen in Fig.\ref{Fig:ChargeGaps}a for the gapless
sector of the phase diagram at $\mathrm{h}^{ext}_{c1} <
\mathrm{h}^{ext} < \mathrm{h}^{ext}_{c2}$ the two lowest states
cross each other $N/2$ times, that the first crossing occurs at
point $\mathrm{h}^{ext}_{c1}$ and the last crossing occurs at the
point $\mathrm{h}^{ext}_{c2}$. In this region we also observe
numerous additional level crossing between the lowest second and
third eigenstates. These level crossings lead to incommensurate
effects that manifest themselves in the oscillatory behavior of
the spin correlation functions. All crossings disappear at
$\mathrm{h}^{ext} > \mathrm{h}^{ext}_{c2} $ and the correlation
functions do not contain oscillatory terms in this region of the
phase diagram.

In marked contrast with the isotropic case, the similar analysis
of the few lowest levels for an {\em anisotropic} ladder in the
presence of transverse magnetic field reveal a principally
different behaviour.  The energy difference between  the two
lowest levels and the ground state energy as a function of the
magnetic field $\mathrm{h}^{ext}$ has been computed for
$J_{\bot}=3J$ and $J=1.0$ for different ladder lengths
$N=12,16,20$ and different values of the anisotropy parameter
$\Delta=0.3,0.5,0.7$. As an example, in Fig.\ref{Fig:ChargeGaps}b
we have plotted results of this calculations for $\Delta=0.5$. As
it is seen from the figure, the excitation spectrum in this case
is gapfull except at the two critical points
$\mathrm{h}^{ext}_{c1} = 2.3\pm0.1$ and $\mathrm{h}^{ext}_{c2}=
3.0 \pm 0.1$.  In the region of magnetic fields
$\mathrm{h}^{ext}_{c1}< \mathrm{h}^{ext} < \mathrm{h}^{ext}_{c2}$
the two lowest states form a twofold degenerate ground state in
the thermodynamic limit and the spin gap, which appear at
$\mathrm{h}^{ext} > \mathrm{h}^{ext}_{c1}$, first increase vs
external field and after passing a maximum decreases to
vanishe at
$\mathrm{h}^{ext}_{c2}$. At $\mathrm{h}^{ext} >
\mathrm{h}^{ext}_{c2}$ the gap once again opens and, for a
sufficiently large field becomes proportional to
$\mathrm{h}^{ext}$. These results are in good agreement with the results obtained
in the studies of the fully anisotropic antiferromagnetic $XYZ$ chain in a magnetic field
\cite{hogemans}.

To determine the properties of this model in different sectors of the
phase diagram we have implemented the modified Lanczos algorithm
of finite size ladders $(N=12, 16, 20, 24)$ to calculate the
magnetizations $M^{x,y,z}$ and spin structure factors ${\cal
S}^{xx}(q)$, ${\cal S}^{yy}(q)$ and ${\cal S}^{zz}(q)$.


\begin{figure}[tbh]
\centerline{\includegraphics[width=\smallfig]{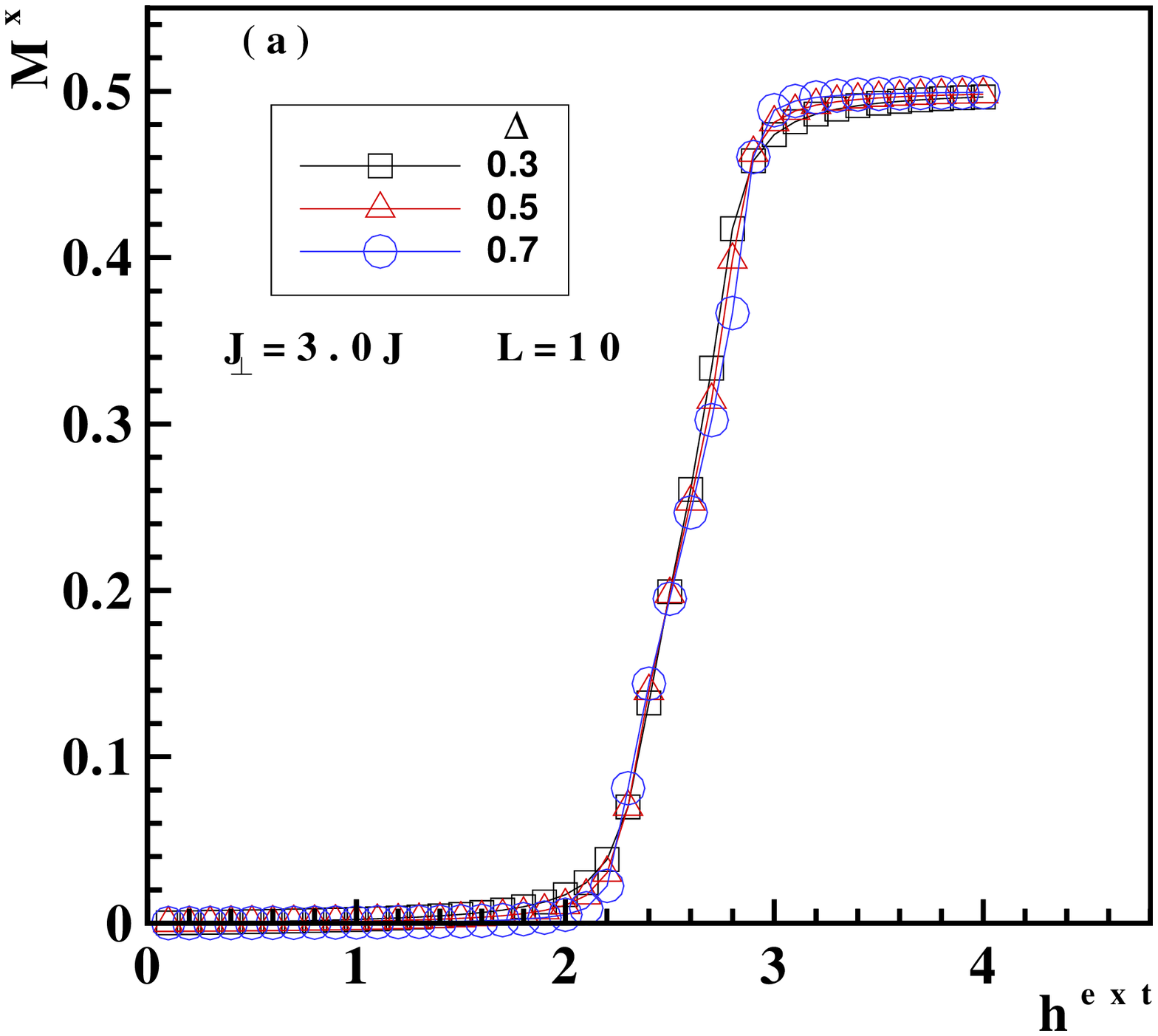}
\includegraphics[width=\smallfig]{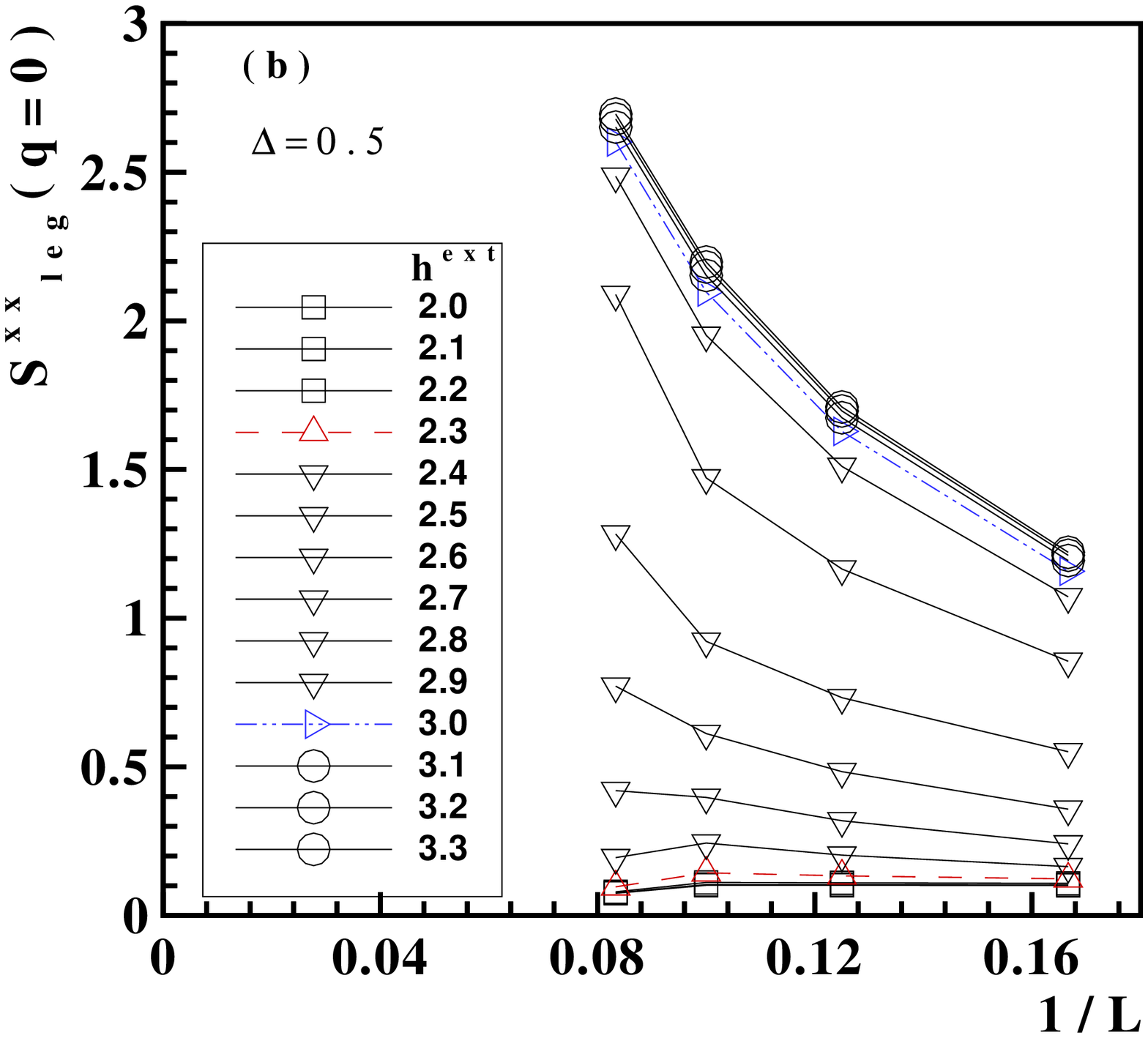}}
\caption[DIAGONAL]{a. The transverse magnetization $M_{x}$ as a
function of applied field $\mathrm{h}^{ext}$ for $L=10$ ladder
with $J_{\bot}=3.0 J$ and for different values of the anisotropy
parameter $\Delta=0.3, 0.5, 0.7$.
b. The intraleg spin structure factor in the x direction at $q=0$
for $J_{\bot}=3.0 J$ and $\Delta=0.5$ plotted as a function of the
inverse ladder length $L$ ( $L=N/2$ ) for different strength of
the applied field $h^{ext}$. The dashed line which corresponds to
the $h^{ext}=h^{ext}_{c1}=2.3J$ marks the transition from the RS
phase to the intermediate phase with finite magnetization along
the field and the dashed-dotted line corresponds to
$h^{ext}=h^{ext}_{c2}=3.0J$ and marks the transition into the
phase with full magnetization along the field.}
\label{Fig:Mx-SFxx}
\end{figure}


In Fig.\ref{Fig:Mx-SFxx}a we have plotted the magnetization along
the applied transverse magnetic field, $M^{x}$ vs
$\mathrm{h}^{ext}$ for  $N=20$, $J_{\bot}=3J$ and for different
values of the  anisotropy parameter $\Delta=0.3,0.5,0.7$. Due to
the profound effect of quantum fluctuations the transverse
magnetization remains small but finite for
$0<\mathrm{h}^{ext}<\mathrm{h}^{ext}_{c1}$ and reaches zero at
$\mathrm{h}^{ext}=0$. This is in agreement with results obtained
within the weak-coupling continuum-limit bosonization treatment
\cite{VJM_2} and with the results for the magnetization obtained
in the case of fully anisotropic  $XYZ$ chain \cite{hogemans}. For
$\mathrm{h}^{ext} > \mathrm{h}^{ext}_{c1}$ the magnetization
increases linearly with increasing field once again in complete
agreement with the predictions of  bosonization analysis
\cite{VJM_2}. This behavior is in agreement with expectations,
based on the general statement that in the gapped rung-singlet
phase, magnetization along the applied field appears only at a
finite critical value of the magnetic field equal to the spin gap
[13-16]. However, in finite systems we do not observe a sharp
transition at this point ($\mathrm{h}^{ext}_{c1}$) or close to the
saturation value which happens at $\mathrm{h}^{ext}
> \mathrm{h}^{ext}_{c2}$.  Magnetization along the directions
perpendicular to the applied field remains small, but finite.
However, as we will show below, for $\mathrm{h}^{ext}_{c1} <
\mathrm{h}^{ext} < \mathrm{h}^{ext}_{c2}$ the intraleg
magnetization along the "y" direction is finite and shows the
long-range ferromagnetic order.

An additional insight into the nature of different phases can be obtained
by studying the spin-spin correlation functions. In particular we
study the magnetic field dependence of the different spin structure
factors. We have calculated the intraleg spin
structure factors defined by
\begin{eqnarray}
{\cal S}_{leg}^{\alpha\beta}(q)=\frac{1}{N} \sum_{n,r} \langle
\,0| \,s_{j,n}^{\alpha}s_{j,n+r}^{\beta}\,|0\,\rangle \, e^{i q r}
\, .
\end{eqnarray}

The field dependence of the intraleg spin structure factor ${\cal
S}_{leg}^{xx}(q=0)$ is qualitatively the same as of the transverse
magnetization $M_{x}$. In Fig.\ref{Fig:Mx-SFxx}b we have plotted
the intraleg spin structure factor ${\cal S}_{leg}^{xx}(q=0)$ for
different strength of the applied magnetic field chosen in the
vicinity of $h^{ext}_{c1,c2}$ as a function of the ladder length
$L$. As it is  seen from this figure the dashed line which
corresponds to the $h^{ext}=h^{ext}_{c1}=2.3\pm0.1$ marks the
transition from the RS phase to the intermediate phase with finite
magnetization along the field and the dash-dotted line which
corresponds to $h^{ext}=h^{ext}_{c2}=3.0\pm0.1$  marks the
transition into the phase with saturate magnetization along the
field.

\begin{figure}[tbp]
\centerline{\includegraphics[width=0.5\columnwidth]{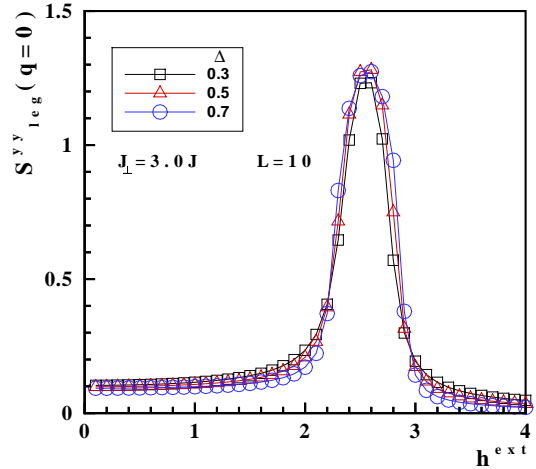}}
\caption{The intraleg spin structure factor ${\cal
S}_{leg}^{yy}(q=0)$ plotted as a function of the applied field
$\mathrm{h}^{ext}$ for $L=10$ ladder with $J_{\bot}=3.0J$ and for
different values of the anisotropy parametere  $\Delta=0.3, 0.5,
0.7$. } \label{Fig:FL_3_Fig_SFyy_D_03-05-07_Jtr_3.eps}
\end{figure}

In Fig.\ref{Fig:FL_3_Fig_SFyy_D_03-05-07_Jtr_3.eps} we have
plotted ${\cal S}_{leg}^{yy}(q=0)$ as a function of the applied
field $\mathrm{h}^{ext}$ for  $L=10$, $J_{\bot}=3J$ and different
values of the anisotropy parameter $\Delta=0.3, 0.5, 0.7$. As it
is clearly seen from this figure, in complete agreement with the
bosonization results \cite{VJM_2}, there is not any long range
ferromagnetic order along the $"y"$ direction at
$\mathrm{h}^{ext}<\mathrm{h}^{ext}_{c1}=2.3\pm0.1$ and
$\mathrm{h}^{ext} > \mathrm{h}^{ext}_{c2} =3.0\pm0.1$. However, in
the intermediate region  $\mathrm{h}^{ext}_{c1}< \mathrm{h}^{ext}
< \mathrm{h}^{ext}_{c2}$, the spins of each leg show a profound
ferromagnetic order in the $"y"$ direction.


\begin{figure}[tbh]
\begin{center}
\centerline{\includegraphics[width=\smallfig]{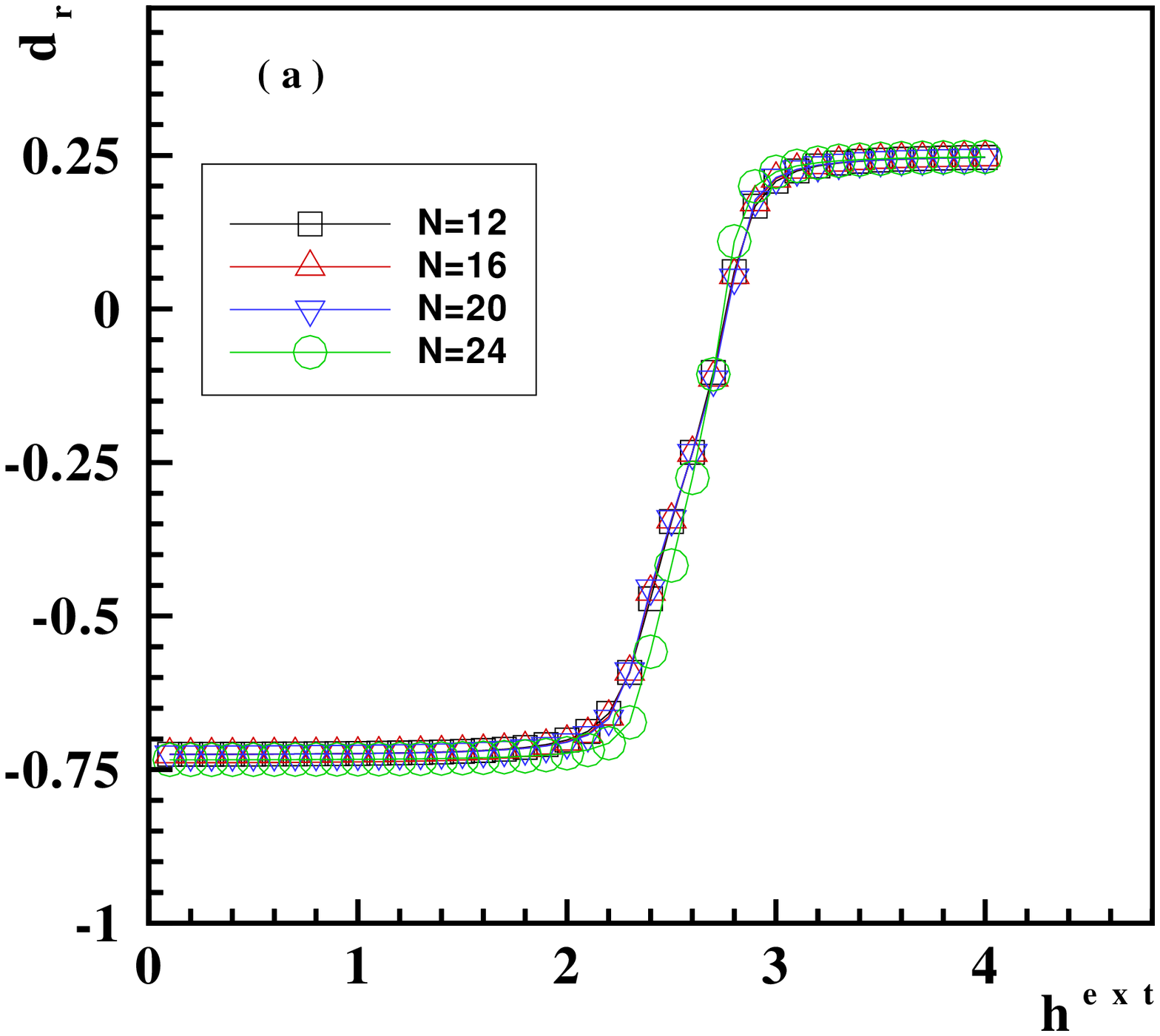}
\includegraphics[width=\smallfig]{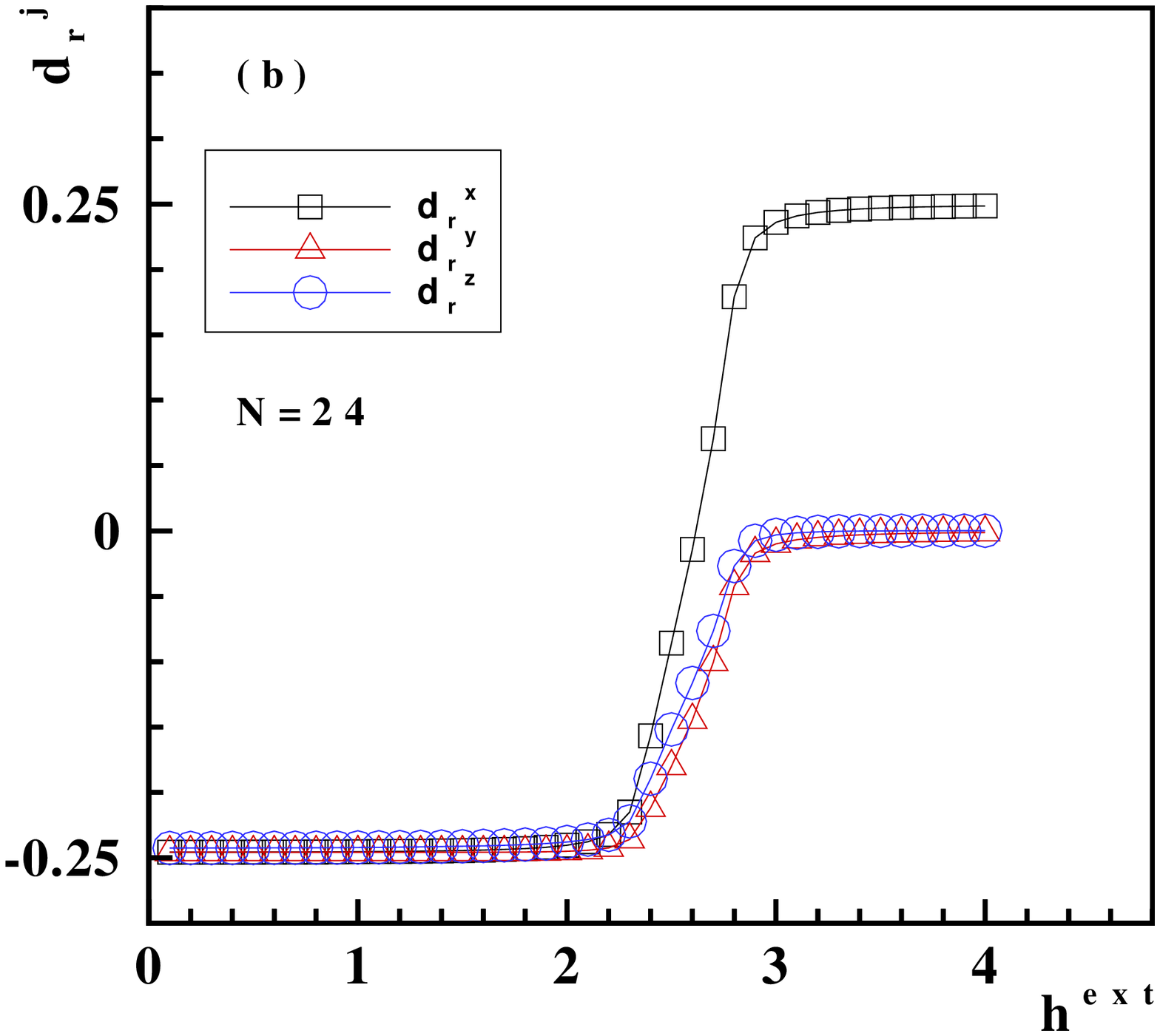}}
\caption[DIAGONAL]{a. The rung-singlet order parameter $d_{r}$ as a function of
the applied magnetic field $h^{ext}$ for the ladder with
$J_{\bot}=3.0J$, $\Delta=0.5$ and for different values of the ladder
length $N=12,16,20,24$.
   b.  The rung-singlet order parameter $d^{x}_{r}$, $d^{y}_{r}$ and
$d^{z}_{r}$ as a function of the applied magnetic field $h^{ext}$ for the ladder with
$J_{\bot}=3.0J$, $\Delta=0.5$ and $N=24$.}
\label{Fig:dr-dxyz}
\end{center}
\end{figure}

To display the magnetic phase diagram of the anisotropic ferromagnetic ladder in
a transverse magnetic field  we have also calculated numerically the {\em intrarung}
spin correlation functions. We have computed the on-rung dimerization order parameter given by
\begin{eqnarray}
d_{r} = \frac{1}{N} \sum_{n} \langle \,0| \,\overrightarrow{s}_{1,n}
\cdot \overrightarrow{s}_{2,n}\,|0\,\rangle \, .
\end{eqnarray}
In Fig.~\ref{Fig:dr-dxyz}a we have plotted $ d_{r} $ as a function
of $\mathrm{h}^{ext}$ for the ladder with $J_{\bot}=3J$,
$\Delta=0.5$ and for different values of the ladder lengths
$N=12,16,20,24$. As it is clearly seen from this figure for $h^{ext} <
h_{c1}^{ext}$, $d_{r}$ is close to $-0.75$) and the ladder is in the
rung singlet phase. For $h^{ext} > h_{c2}^{ext}$, $d_{r}$ is slightly
less then the saturation value $d_{r}\sim 0.25$ and the
ferromagnetic long range order along the $"x"$ axis is present.
Deviation from the saturation values $-0.75$ and $0.25$ is the  result of
 quantum fluctuations. In this case of finite systems and
with chosen values of the rung exchange, the critical magnetic
fields are high which strongly suppresses the quantum
fluctuations. As a result the obtained averages of spin correlation
functions are very close to their classical saturation values. In the
intermediate range of field the system smoothly evaluates from a RS
phase into the FM phase via formation of the stripe-ferromagnetic
order.

To get additional data about the character of the spin ordering in the intermediate SFM phase we have calculated the "$x$","$y$" and "$z$"
components of the rung-singlet order parameter given by
\begin{equation}
d^{j}_{r} = \frac{1}{N} \sum_{n} \langle \,0| \,s^{j}_{1,n}s^{j}_{2,n}\,|0\,\rangle \, , \qquad j=x,y,z\, .
\end{equation}
In Fig.\ref{Fig:dr-dxyz}b we have plotted $d^{x}$, $d^{y}$ and
$d^{z}$  as a function of the applied field $\mathrm{h}^{ext}$ for
$N=24$ ladder with $J_{\bot}=3J$ and for $\Delta=0.5$. As it is
seen from this figure in the RS, at $\mathrm{h}^{ext} <
\mathrm{h}^{ext}_{c1}$ the system clearly shows the $SU(2)$
invariant order within the rung $d^{x}=d^{y}=d^{z}\sim
-0.25$. For $\mathrm{h}^{ext} > \mathrm{h}^{ext}_{c1}$ the $SU(2)$
invariance is broken, $d^{x}$ increases with increasing field
faster then the other two components and soon approaches the positive
values corresponding to the ferromagnetic order in the "$x$"
direction, while $d^{y}=d^{z}$ remain negative indicating the
antiferromagnetic correlations in these components.
This shows the suppressed
ferromagnetic order in the "$z$" direction and profound
ferromagnetic response of the "$y$" direction presents completely
convincing arguments in support of statement, that for
intermediate values of applied field, at $\mathrm{h}^{ext}_{c1}<
\mathrm{h}^{ext} < \mathrm{h}^{ext}_{c2}$ ferromagnetic ordering
in the $"x"$ direction of spin from both legs is accompanied by
antiferromagnetic order in the $"y"$ direction of spins located on
different legs and completely suppressed correlations in the $"z"$
direction. It corresponds to the stripe ferromagnetic
phase. Finally for $\mathrm{h}^{ext} > \mathrm{h}^{ext}_{c2}$  the ordering
in $"y"$ and $"z"$ directions are completely suppressed and the
system shows ferromagnetic order with nominal magnetization per
spin along the applied field.

\section {Conclusions}

We have studied the ground state phase diagram of the two-leg spin ladders with ferromagnetic anisotropic legs in a transverse magnetic field. We have implemented the modified Lanczos method to get the excited state energies as the same accuracy of the
ground state one. Two quantum phase transitions in the ground state
of the system with increasing magnetic field have been identified.
The first transition is the gapped
rung-singlet  to the gapped stripe-ferromagnetic phase. The
second - to the transition from the gapped stripe-ferromagnetic
phase into the fully polarized ferromagnetic phase. This results
are in complete agreement with the results obtained from the
bosonization treatment \cite{VJM_2}.

\section{Acknowledgments}

It is our pleasure to thank D. Cabra, H.-J. Mikeska, E. Pogosyan
and T. Vekua for fruitful discussions. GIJ would like to thank A.
Ferraz for interesting discussions and kind hospitality during his
stay in ICCMP-UnB. He also acknowledges support from the CNPq and
the Ministry of Science and Technology of Brazil and support of
the Georgian National Science Foundation (Grant GNSF/ST06/4-018).


\end{document}